\documentstyle[12pt,epsf]{article}

\renewcommand{\baselinestretch}{1.5}

\newcommand{\alt}{\raisebox{-.5ex}{\scriptsize $\stackrel{\textstyle <}{ \sim}$}}
\newcommand{\agt}{\raisebox{-.5ex}{\scriptsize $\stackrel{\textstyle >}{ \sim}$}}

\begin{document}

\begin{center}

{\bf Internal Forces and the Magnetoconductivity\\
of a Nondegenerate 2D Electron Fluid}

P. Fozooni, P.J. Richardson, and M.J. Lea

{\it Department of Physics, Royal Holloway, University of London, Egham,
Surrey TW20 0EX, England}

M.I. Dykman$^{(a)}$ and C. Fang-Yen$^{(b)}$\cite{byline}

$^{(a)}$ {\it Department of Physics and Astronomy, 
Michigan State University, MI 48824\\
$^{(b)}$ Department of Physics, Stanford University, Stanford, CA 94305}

A. Blackburn

{\it Department of Electronics and Computer Science, 
University of Southampton, Southampton SO17 1BJ, England}

\end{center}

\renewcommand{\baselinestretch}{1.0}
\begin{quote}
{\small The forces on individual electrons in an unscreened nondegenerate electron
fluid, due to electron density fluctuations, have been
calculated using Monte Carlo simulations and determined experimentally
over a broad range of the plasma parameter $\Gamma$.  The experimental
results
are obtained from the magnetoconductivity \(\sigma(B)\) measured for
electrons on liquid helium below 1 K for $B \leq 8$ T.  The magnitude and
density dependence of \(\sigma(B)\) are explained by the many-electron
theory of magnetotransport. The internal electric fields found from the 
experiments are in excellent agreement with the simulations.}
\end{quote}


\renewcommand{\baselinestretch}{1.5}

Electrons above the surface of superfluid helium often form a
two-dimensional (2D) {\it normal fluid} \cite{2D}:
for characteristic electron densities \(n\sim 10^{12}\) m$^{-2}$ the
interelectron distance
\(\sim n^{-1/2}\) exceeds the de Broglie wavelength, 
and the system is nondegenerate; at the same time, the ratio of the
characteristic Coulomb energy of electron-electron interaction to the
kinetic energy, the plasma parameter $\Gamma = e^2(\pi n)^{1/2}/4\pi
\epsilon_0 kT$ is large, and therefore there is
short-range order in the electron system \cite{Gann}. For $\Gamma >
127$ (low $T$) electrons form a 2D crystal \cite{Wigner}. Mean-field
effects such as long-wavelength plasma oscillations are well
understood for a normal 2D electron liquid \cite{plasma} but much less
is known about the detailed behavior of individual electrons. An
experimental probe is required while, on the theoretical side, the
problem is complicated by the absence of ``good" quasiparticles.

In this Letter, we present data from Monte Carlo simulations and 
experimental measurements of an important but unexplored
characteristic of a 2D electron fluid, the internal electric field ${\bf
E}_{\rm f}$ that drives an electron as a result of its interaction
with other electrons. Unlike the long-wavelength fluctuational fields
known in plasma physics \cite{Krall}, the field ${\bf E}_{\rm f}$,
although also of fluctuational origin, determines the force
driving an {\it individual particle}, and is not described by the theory
\cite{Krall}.  The force on a particle is an important  dynamical
characteristic of a system. In ``conventional'' fluids with short-range
interatomic interaction, the forces have been a subject of extensive research
\cite{Croxton}. The mean square force remains
finite in a fluid (though the mean square displacement diverges), but
it would be expected to display singular behavior at the
liquid-crystal transition.

A special significance of the field ${\bf E}_{\rm f}$ for 
a nondegenerate 2D electron system stems from the fact that, 
over a broad range of parameters, it strongly affects
magnetotransport. In the 
single-particle approximation the electron energy spectrum in a 
magnetic field $B$ perpendicular to the electron layer consists 
of discrete Landau levels, separation $\hbar \omega_c$ 
($\omega_c = eB/m$ is the cyclotron frequency). 
The centers of the cyclotron orbits move only because of the random potential
of the scatterers. Therefore scattering is always strong, and 
the pattern of transport is very different from the standard 
Drude picture which applies for weak coupling at $B = 0$ and in 
which scattering events are short and well separated in time.

The field ${\bf E}_{\rm f}$ causes the centers of the cyclotron
orbits to drift at a velocity $E_{\rm f}/B$, and may
\lq\lq restore" the weak coupling picture. This effect 
was first discussed and investigated \cite{Khazan} 
for transport in the quantum limit $\hbar \omega_c/kT 
\gg 1$. It was shown later \cite{PRL93,Lea94} that the field ${\bf E}_{\rm 
f}$ also dramatically affects the magnetoconductivity $\sigma (B)$ in
classically strong magnetic fields, $\mu B \gg 1, \, \hbar \omega_c/kT
< 1$ ($\mu$ is the zero-field electron mobility), for vapor-atom
scattering above 1 K.  In particular, it restores the Drude-type
behavior of $\sigma (B)$ for comparatively small $B$ which has been known
experimentally since \cite{Iye}. Cyclotron resonance measurements
\cite{CR} have also demonstrated the importance of Coulomb
interactions in this system. Several other mechanisms have been
proposed to explain previous measurements of $\sigma (B)$ at higher
$B$ \cite{Kovdrya}. 

Our new measurements of $\sigma (B)$ have been done below 1 K, in the
ripplon scattering regime. In this range the mobility $\mu$ is
extremely high ($\alt 2000\, {\rm m}^2/{\rm V s}$), enabling
{\it quantitative} characterization of ${\bf E}_{\rm f}$ as
a function of density and temperature.

We consider the distribution of  ${\bf E}_{\rm f}$ for a 
classical normal liquid; the results also apply for quantizing 
magnetic fields provided the motion of the centers of the cyclotron 
orbits is semiclassical. Since fluctuations in the system are 
thermal, and the field is due to electron-electron 
interaction, the scale for ${\bf E}_{\rm f}$ is given by 
the characteristic field $E_0$: 

\begin{equation}
\langle {\bf E}_{\rm f}^2\rangle = F(\Gamma)E_0^2, \; E_0 =
\left(kTn^{3/2}/ 4\pi\epsilon_0\right)^{1/2}. 
\end{equation}

The scaled dimensionless mean square field $F(\Gamma)$ can be
easily found for large $\Gamma$ (low $T$)
in the 2D crystal phase
\cite{JPhC82}. Here, the force on an electron $e{\bf E}_{\rm f}$
arises because of the displacement of electrons from their lattice
sites ${\bf R}_n$. In the harmonic approximation the force is linear in the
displacements and has a Gaussian distribution. The function $F$
incorporates contributions from both transverse and longitudinal modes
of the crystal, and $F(\Gamma) \approx 8.91$, independent of $\Gamma$.

In the opposite limit of a nearly ideal plasma, $\Gamma \ll 1$, 
the major contribution to the force on an electron comes from 
pair collisions, which gives $F(\Gamma) \approx 2\pi^{3/2}/\Gamma$.  

In the most interesting range of the electron liquid and the melting
transition, the function $F(\Gamma)$ was obtained from Monte Carlo
simulations. We used the Metropolis algorithm and the Ewald summation
technique as in \cite{Gann}(a), with periodic boundary conditions. The
field on an electron was evaluated as the gradient of the potential in
which the electron was moving.

The results for ${\bf E}_{\rm f}$ for the number of particles $N =
196, 324$ are very close to each other. In the range $\Gamma > 30$
the probability density distribution $\rho$ of the field components
$E_x,\,E_y$ is close to Gaussian. The functions $F(\Gamma)$ and $\rho$
are plotted in Fig.~1. The scaling function $F(\Gamma)$ decreases
nearly monotonically with increasing $\Gamma$.  However, remarkably,
its variation is {\it very small} in the range $\Gamma \agt 10$, although the
structure of the system changes dramatically, from a crystal to a
liquid with a correlation length of twice the mean electronic
separation. The function $F(\Gamma)$ has a smeared singularity at the
melting point.  Detailed discussion will be given elsewhere
\cite{tobepublished}.

The magnetoconductivity of 2D electrons above superfluid helium was
measured using 4 mm diameter Corbino disk electrodes (see Fig.~2) 
100 $\mu{\rm m}$ beneath the electrons \cite{Lea94}. 
Free electrons were held over the central drive
electrode {\bf A}, a ring electrode {\bf E}, and 
receiving electrodes {\bf B1}, {\bf B2} and {\bf B3} surrounded by a
planar guard {\bf G}.  An ac voltage $V_0$
(typically 10 mV) at a frequency up to 10 kHz was applied to
electrode {\bf A} and the ac currents $I$ to the electrodes {\bf B}
measured. For a perfect conductor the phase of the capacitively
coupled current $I$ is $\pi /2$ with respect to $V_0$. The phase shift
$\phi(B)$ away from $\pi/2$ was measured for perpendicular magnetic
fields $B$ {\raisebox{.4ex}{\scriptsize $\leq$}} 8 T for electron densities $0.5 \times 10^{12}\alt
n\alt 2\times 10^{12}$ m$^{-2}$ at temperatures $0.6 \alt T \alt 0.9$
K in the fluid phase. The phase shift $\phi(B)$ is proportional to 
$\sigma^{-1} (B)$ for $\phi \alt 0.3$ rad, while for larger phase 
shifts the theoretical response function was used. The density $n$ 
was determined from the -ve dc
bias voltage on electrode {\bf E} required to cut off the current
between electrodes {\bf A} and {\bf B}.

Figs.~2 and 3 show the measured magnetoconductivity for several
densities and temperatures. In low fields, $B < 0.5$ T, the data
accurately follow the simple Drude-like result, even for values of
$\mu B$ as large as 500:

\begin{equation}
\sigma (B) = {\sigma(0)\over 1+(\mu B)^2},\quad 
{ne\over\mu\sigma (B)}\approx B^2 \;\, {\rm for} \,\; \mu B \gg 1.
\end{equation}

\noindent
The electron mobility was determined from the $B^2$ dependence of
$\sigma^{-1}(B)$,
as a function of density and temperature, and is in excellent
agreement with previous $B = 0$ measurements by Mehrotra {\it et al.}
\cite{Mehrotra}. The measured mobilities from 0.6 to 0.9 K are close to the
theoretical values for a classical strongly correlated electron liquid
\cite{tobepublished}, with scattering by both ripplons and $^4$He vapor atoms 
taken into account \cite{Saitoh}. The mobility is slightly
density-dependent, primarily because the electric field that presses
electrons against the helium surface increases with $n$, and therefore
so does the electron-ripplon coupling. The data is plotted as
$ne/\mu\sigma (B)$ vs. $ B$ (Fig.~2) or $B^2$ (Fig.~3) using the
experimental values of $\mu$ for each $n$ and $T$.  For $B < 0.5$ T the
data lie on the {\it universal} line, $ne/\mu\sigma (B) = B^2$ (line
{\bf a} in Figs.  2 and 3).

At this point we should stress that the simple-minded Drude model (2)
effectively applies {\it because of the internal electric fields}
\cite{PRL93,tobepublished}. To understand the effect qualitatively, we
notice that the many-electron system transfers its momentum to
short-range scatterers via individual electron-scatterer collisions.
In a certain range of the parameters, all that an electron ``knows''
about other electrons during a collision is the field ${\bf E}_{\rm
f}$, and this field is time-independent if the collision is short
enough.  Then the Einstein relation for the conductivity applies:

\begin{equation}
\sigma (B) = ne^2L^2\tau^{-1}(B)/ kT,
\end{equation}

\noindent
where $L \equiv L(B)$ is the diffusion length and $\tau^{-1}(B)$ is
the relaxation rate. For $\mu B \gg 1$ the diffusion length is given
by the mean radius $\bar R$ of the cyclotron orbit, $L^2 = \bar R^2/2 = 
(\hbar/2eB)(2\bar n +1)$, with $\bar n = 1/[\exp(\hbar\omega_c/kT)-1]$.

It is the relaxation rate in (3) that is primarily affected by
internal electric fields. For $kT\gg eE_{\rm f}
\lambda_T$\raisebox{.6ex}{\hspace{-11pt}--} $\,>
\hbar\omega_c$ 
($\lambda_T$\raisebox{.6ex}{\hspace{-11pt}--} 
$\, = \hbar/(2mkT)^{1/2}$) the field ${\bf
E}_{\rm f}$ smears out the Landau levels and thus {\it eliminates} the
effects of their discreteness.  Therefore $\tau^{-1}(B) =
\tau^{-1}(0)$, and the $B$-dependence of $\sigma$ is given by that of
$L^2 =mkT/e^2B^2\propto B^{-2}$, i.e. the many-electron theory
gives $en/\mu\sigma$ independent of $n$ as observed, cf.
the solid line {\bf a} in Figs.~2 and 3.  In contrast, in the
single-electron picture the relaxation rate is increased by the density
of states enhancement factor $\omega_c\tau(B)$, since the states in
the energy strip $\hbar\omega_c$ are ``compressed'' down to the Landau
level collision width $\hbar\tau^{-1}(B)$. The total
magnetoconductivity from the self-consistent Born approximation (SCBA)
for ripplon
\cite{Saitoh} and gas-atom scattering \cite{Ando} and their 
self-consistent combination $\sigma_{\rm s} = (\sigma_{\rm rs}^2 +
\sigma_{\rm
gs}^2)^{1/2}$ is plotted (line {\bf b}) in Fig. 2 for $n = 0.55\times
10^{12}$m$^{-2}$ and $T =$ 0.7 K.  At 2 Tesla, the
SCBA overestimates $\sigma(B)$ by an order of magnitude.

A distinctive feature of the classical $\sigma(B)$ is {\it saturation}
with increasing $B$. This arises because an electron in crossed ${\bf
E}_{\rm f}$ and ${\bf B}$ fields moves along a spiral with a step
$\sim 2\pi E_{\rm f}/B\omega_c$. The number of times it encounters a
short range scatterer in the classically strong field $B$ is then
$N_{\rm enc}
\sim \lambda_T$\raisebox{.6ex}{\hspace{-11pt}--} 
$B\omega_c/2\pi E_{\rm f}$ ($\lambda_T$\raisebox{.6ex}{\hspace{-11pt}--} $\,$ is the uncertainty
in the position of an electron). For $N_{\rm enc} > 1$ one would
expect the scattering rate, and thus $\sigma (B)$, to increase by
a factor $\sim N_{\rm enc}
\propto B^2$, and we find (cf. Eq.(2))

\begin{equation}
{ne\over\mu\sigma(B)} \approx \pi B_0^2, \quad B_0^2 =
\left({2m^3kT\over e^2\hbar^2}\right)^{1/2}\langle 
{\bf E}_{\rm f}^2\rangle ^{1/2}.
\end{equation}

\noindent
The saturation of $ne/\mu\sigma(B)$ is clearly seen in Figs.~2 and 3
for $B > 1.0$ T. The limiting value of $ne/\mu\sigma(B)$ increases with
density (Fig.~2) and temperature (Fig.~3) and is directly proportional
to the rms internal electric field, from Eq.(4). It is this which enables the
internal field to be determined experimentally. 

For $B \agt 1$ T quantum effects become substantial. 
For $\hbar\omega_c \gg kT$ the diffusion
length $L \approx (\hbar/2eB)^{1/2}$, and also $N_{\rm enc}
\sim (\hbar/eB)^{1/2}B\omega_c/2\pi E_{\rm f}$, so that $ne/\mu\sigma(B)
\propto B^{-1/2}$ decreases with increasing $B$. For higher 
$B \agt 5$ T (depending on $n$ and $T$) the duration of a collision
$\tau_{\rm col} \agt \tau(B)$ and this theory no longer applies.

Eq.(4) is written \cite{PRL93} for short-range scatterers (such as
helium vapor atoms). In the case of scattering by ripplons, an extra
factor arises which is numerically close to 1 for the
experimental conditions here. Calculations of the magnetoconductivity 
for gas-atom and ripplon scattering were made using the full 
semiclassical many-electron theory in the range $\hbar\omega_c \agt kT$ 
\cite{tobepublished}, with values of $\langle {\bf E}_{\rm
f}^2\rangle$ taken from Fig.~1. The results for the total many-electron
magnetoconductivity $\sigma_{\rm m} = \sigma_{\rm rm} + \sigma_{\rm
gm}$ are shown in Fig.~2 (lines {\bf c}, {\bf d} and {\bf e},
increasing $n$) and Fig.~3 (lines {\bf b}, {\bf c}, {\bf d } and
{\bf e}, increasing $T$), and show satisfactory agreement with the
experiments for $\tau_{\rm col}\ll \tau(B)$ (the extrapolation of the
theoretical curves to the range $\tau_{\rm col}\sim \tau(B)$ is shown
dashed).

Conversely, the measured $\sigma^{-1}(B)$ at $B = 2$ T was used to
obtain experimental values of the internal electric fields.  This
value of $B$ is within the range of the applicability of the theory
($kT > eE_{\rm f}(\hbar/eB)^{1/2}$; $\hbar\omega_c > kT$) and is also
far from the Drude region and the region where collisional level
broadening affects the scattering. The experimental values of
$\langle {\bf E}_{\rm f}^2\rangle^{1/2}$ vs. $E_0$ are shown in Fig.~4
(we renormalized $E_0$ in Eq.(1) to allow for the dielectric constant
of liquid helium). The points come from over 40 combinations of
density and temperature between 0.6 and 0.9 K where the conductivity
varies by more than an order of magnitude; no adjustable parameters
have been used, and so the spread of the points may be considered to
be within reasonable limits.  Within the errors the measured field
is $\propto E_0$, with a constant of proportionality $\nu = 3.11 \pm
0.10$. This can be compared with $F^{1/2} = 3.07 \pm 0.03$ from the
Monte Carlo simulations for the range $20 < \Gamma < 70$ covered by
the experiments. A slight decrease of $\nu$ with decreasing $T$ seen in
Fig.~4 lies within the errors. 

In conclusion, we have both computed and measured the internal
electric fields in a nondegenerate 2D electron fluid, and the results
are in excellent agreement. We show that, over a broad
range of parameters, the magnitude and density dependence of the
magnetoconductivity $\sigma(B)$ of electrons on helium are determined
by many-electron effects and can be understood qualitatively in
terms of electron diffusion controlled by a fluctuational internal
electric field.

We thank R. van der Heijden, P. Sommerfeld and O.Tress for useful
discussions; the EPSRC (UK) for a Research Grant and for a Studentship
(for PJR); the EU for support under contract CHRXCT 930374; A.K.
Betts, F. Greenough and J. Taylor for technical assistance; D. Murphy,
A. Jury, and the staff of the Southampton University Microelectronics
Center and the lithography unit of the Rutherford Appleton Laboratory.

\newpage
\noindent
\underline{Figure Captions}

\begin{itemize}
\item[Fig.~1.] The scaled mean square field 
$F(\Gamma)$ from Monte Carlo calculations. The asymptotic value of $F$
for a harmonic Wigner crystal is shown  dashed. Inset: the field
component distribution.

\item[Fig. 2.] $ne/\mu\sigma(B)$ versus $B$ for $n =$ 0.55 ({\large
$\circ$}), 0.88 ({\small $\bigtriangledown$}) and 1.89 ({\small
$\Box$}) $\times 10^{12} \,{\rm m}^{-2}$ at 0.7 K.  The mobility $\mu =$
985, 830 and $520\pm 20 {\rm m}^2/{\rm Vs}$, respectively.
Solid lines {\bf a} (low $B$) and {\bf c} -- {\bf e} (high
$B$) show many-electron calculations.  Inset: Corbino
electrode geometry.

\item[Fig. 3.] $ne/\mu\sigma(B)$ versus $B^2$ for $T =$ 0.6 ({\small
$\Diamond$}), 0.7 ({\small $\Box$}), 0.8 ({\small $\bigtriangledown$})
and 0.9 K({\large $\circ$}).  The mobility $\mu =$ 755, 620, 430 and
250 $\pm 15 {\rm m}^2/{\rm Vs}$, respectively. Lines {\bf a} (low $B$) 
and {\bf b} -- {\bf e} (high $B$) 
show many-electron calculations. 

\item[Fig. 4.] Values of the internal field $\langle {\bf E}_{\rm
f}^2\rangle^{1/2}$ versus $E_0$ at 0.6 ({\small $\Diamond$}), 0.7
({\small $\Box$}), 0.8 ({\small $\bigtriangledown$}) and 0.9 K({\large
$\circ$}). Solid line shows the best linear fit with $\langle {\bf
E}_{\rm f}^2\rangle^{1/2} = 3.11 E_0$.

\end{itemize}

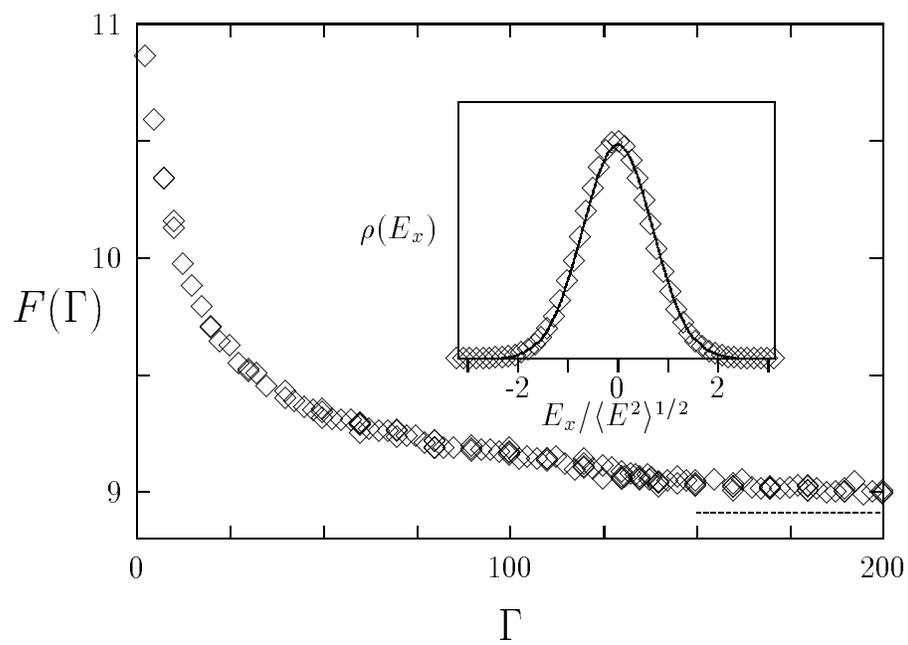
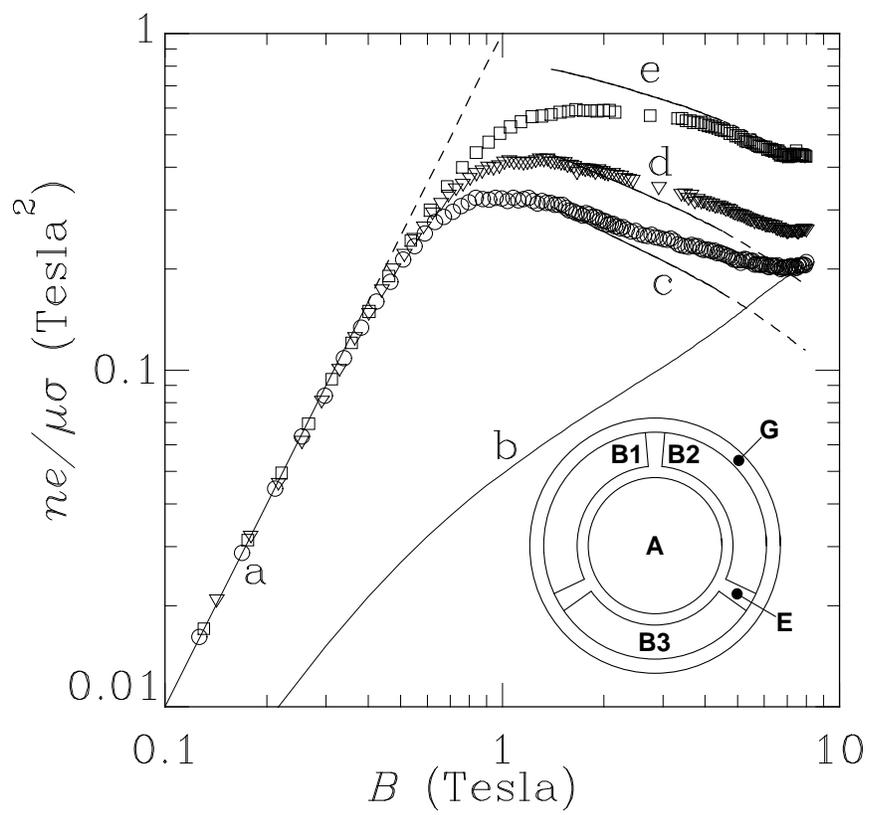
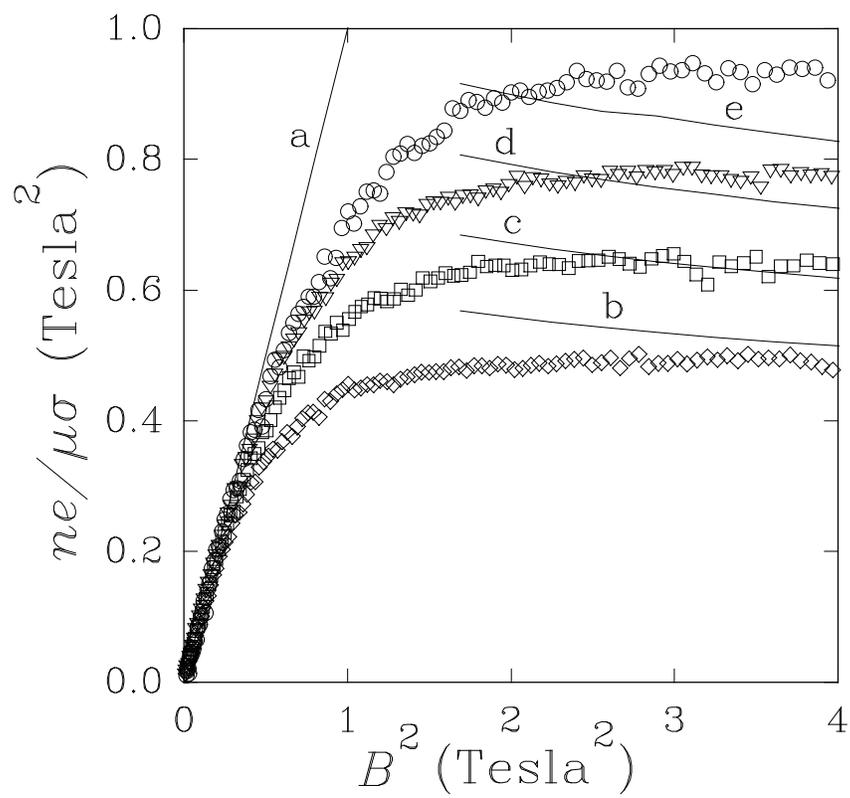
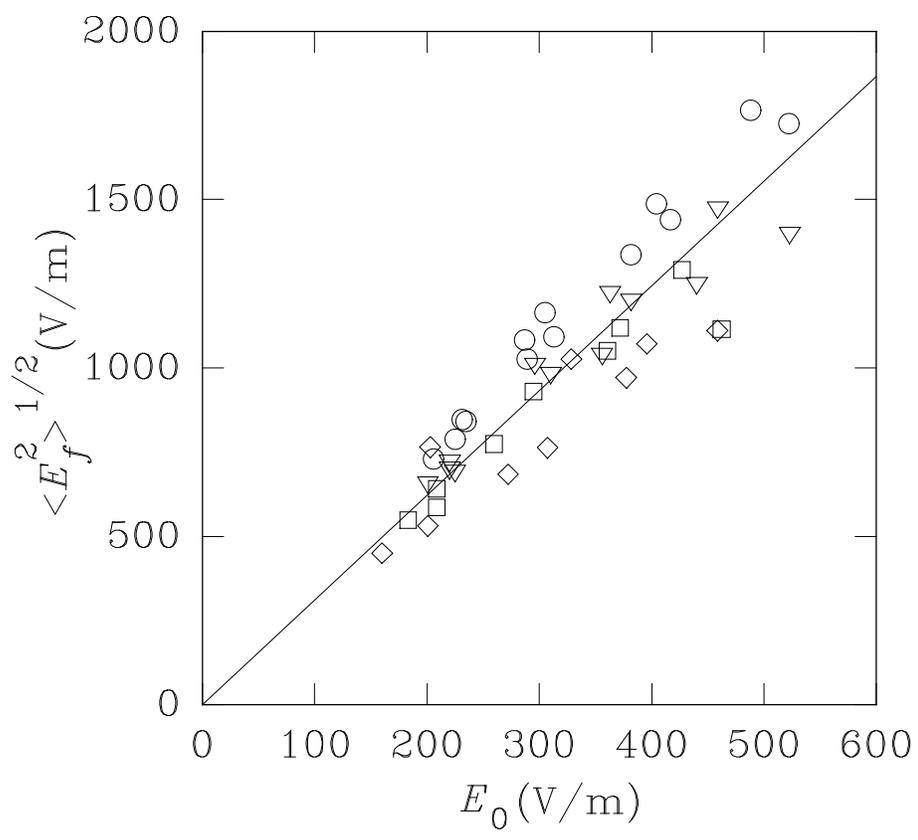
\end{document}